\documentclass[aps,pre,twocolumn,showpacs,amssymb]{revtex4}
\usepackage{bm}

\begin{document}
\preprint{APS}

{\normalfont\small (version 31 August 2015)}

\title{Causal non-locality can arise from constrained replication
}

\author{J. H. van Hateren}
\affiliation{
Johann Bernouilli Institute for Mathematics and Computer Science, University of Groningen, 
Groningen, The Netherlands; j.h.van.hateren@rug.nl
}

\begin{abstract}  
The fundamental theories of physics are local theories, depending on local interactions of local variables. It is not clear if and how strictly local theories can produce non-local variables that have causal effectiveness. Yet, non-local effectiveness appears to exist, such as in the form of memory (non-locality through time) and causally effective spatial structures (non-locality through space). Here it is shown, by construction, how such non-locality can be produced from elementary components: non-isolated systems, multiplicative noise, self-replication, and elimination. A theory is derived that explains how causal non-locality can arise from strictly local interactions.
\end{abstract}

\maketitle

\section{Introduction}

The theories that form the foundation of physics, quantum field theory and general relativity, are local theories~\cite{wil99}. They describe the evolution of local field variables in terms of local interactions in space-time. Such locality is consistent with the empirical facts that physical systems flow contiguously through time and that causal influences cannot travel faster than the speed of light. Nevertheless, local theories are often formulated as non-local ones with non-local variables, if that is convenient for understanding and calculation. For example, finding the dynamics of a system from the principle of least action requires non-local trajectories. Similarly, Maxwell's equations in local, differential form, \textit{e.g.} $\bm{\nabla\cdot}\mathbf{E} = \rho/\epsilon_0$, can be formulated in non-local, integral form, \textit{e.g.} $\oint_{S} \mathbf{E}\bm{\cdot}\mathrm{d}\mathbf{a} = \int_{V} \rho\,\mathrm{d}V /\epsilon_0$. Whereas the first form is purely defined locally, the second form equates non-local quantities obtained by integrating over a non-local surface and a non-local volume.

\vskip-0.1ex
Although non-local formulations are fully equivalent, mathematically, to the corresponding local ones, they are different in the way they map formalism to physical reality. Physical reality is taken to arise from local interactions. Therefore, only local variables are causally effective in the sense that they refer to quantities directly involved in interactions that produce change. In contrast, quantities denoted by non-local variables do not directly interact. They are not directly causally effective themselves. Non-local theories using non-local variables, such as volume and entropy, are often the most natural way to understand a system. But they are taken to be completely explainable from a combination of local causal interactions, at least in principle.

However, there are clear cases, particularly in the realm of life and technology, where non-local variables do seem to have direct causal effectiveness. For example, memory in the form of DNA is a causal factor that appears to act non-locally through time, a spider's web is a non-local spatial structure with causal effectiveness, and also the cylinder and piston of a steam engine only work because of their highly specific spatial structure. The question then arises how non-local variables or structures can get causal effectiveness if all foundational theories are strictly local. Locality seems like a conserved property. In a complex system the interactions may become complex and may strongly vary across space and time, but those interactions would still be local. Yet, in this article I show, by construction, that non-locality with causal effectiveness can indeed arise from local interactions. Local interactions are given in terms of local variables or in terms of non-local variables that are completely defined by a combination of local causal interactions. Such a defining combination does not exist if a non-local variable has causal effectiveness of its own. 

Before proceeding, a disclaimer is necessary. Non-locality is also studied in the context of quantum entanglement and Bell's theorem. But such non-locality concerns correlation rather than causation, and the correlations are fully explained by a local theory~\cite{eng13}. Quantum non-locality is not the topic of this article.

The construction explained below is simplified as much as possible. It should be seen as a mere proof of concept, a stylized version of more elaborate actual systems. The construction proceeds through the following steps. It assumes a population of non-isolated systems that are perturbed by external disturbances. The systems have a limited lifetime and are autocatalytic, that is, can replicate. Replication rates differ between different types of systems, which means that systems with quickly increasing rates will dominate the population. How strongly external disturbances can perturb each system is assumed to depend on the system's structure and momentary state. The form of this dependence that is optimal for replication is derived. This form turns out to depend in a simple way on the replication rate itself. Systems will therefore maximize their abundance in the population if they use an approximation of this rate for modulating their variability. Whereas the real replication rate is a non-local variable without direct causal effectiveness within a system, the approximated replication rate has causal effectiveness through local interactions within that system. In effect, the coupling of these rates provides a non-local variable with causal effectiveness. The next section derives these results in detail.

\section{Theory}

We assume non-isolated systems with a dynamical structure $s$. The systems are capable of self-replication. Systems have a small probability per unit of time to change structure as $s\rightarrow s'$, with $s'$ a small random variation on $s$. The structural space through which $s$ can move is undefined. Systems have a typical lifetime $\tau$ and a time-varying growth rate $k_s(t)$, with their number $n_s(t)$ given by
\begin{equation} \label{eq:dnt}
	\mathrm{d}n_s/\mathrm{d}t = k_s(t) n_s(t),
\end{equation}
with $n_s\geq0$; when $n_s=0$, systems of type $s$ have become extinct. Equation~(\ref{eq:dnt}) produces exponential growth when $k_s(t) > 0$, exponential decline when $k_s(t) < 0$, and stable numbers when $k_s(t) = 0$. The growth rate is assumed to depend on the distance between two real-valued scalars, $E(t)$ and $x_s(t)$. Here $E(t)$ is an environmental variable (written as $E_t$ below), and $x_s(t)$ a state variable of the system. Then
\begin{equation} \label{eq:ki}
	k_s(x_s, t) = k_s(x_s - E_t),
\end{equation}
with $k_s$ maximal at $x_s=E_t$ and monotonically decreasing to $-1/\tau$ for large $|x_s - E_t|$. The latter corresponds to exponential decline when there is no replication. The growth rate thus depends on how well the system state matches the environment. Unlimited growth is prevented by letting $k_s$ decrease uniformly for all systems such that the total number of systems $N(t)=\sum_s{n_s}$ is constrained to a given constant $N_0$. $N_0$ can be thought to depend on a limited availability of raw materials, free energy, and space. Then $N(t)=N_0$ yields
\begin{equation} \label{eq:dN}
	\mathrm{d}N(t)/\mathrm{d}t = \sum_s{\mathrm{d}n_s/dt} = \sum_s{k_s(t) n_s(t)} = 0.
\end{equation}
Because $n_s(t)>0$ for all systems that have not become extinct, the rightmost equality implies that $k_s(t)$ must vary around zero, on average. Variations in $E_t$ and the introduction of new variants $s$ will occasionally drive $k_s$ downwards. Systems that can recover quickly from such decreases by having a large $\mathrm{d}k_s/\mathrm{d}t$ will then gradually replace systems with smaller $\mathrm{d}k_s/\mathrm{d}t$. Systems can therefore maximize the likelihood that their type $s$ persists by maximizing $\mathrm{d}k_s(t)/\mathrm{d}t$ rather than $k_s(t)$ itself. This maximization must be constrained by the condition that systems $s$ do not become extinct. Below we will derive conditions for such a constrained maximization.

The environmental variable $E_t$ is assumed to vary unpredictably, with power distributed across many time scales, both smaller and larger than $\tau$~\cite{bel10,hat15}. It can be thought to arise from a random walk-like process, but band-limited and with a non-uniform, typically power-law spectral density (like coloured noise,~\cite{han95}; $E_t$ is not assumed to be zero-mean, but its time derivative is). The process generating $E_t$ is taken to be independent of the other random processes, in particular the process generating new systems $s$ including their $\sigma_s$ (see below) and the Wiener process $W_t$ (see below). Independence is interpreted here as the assumption that the processes are in no way causally related.

The state variable $x_s$ of a system $s$ is assumed to evolve according to a random walk with state- and time-dependent drift and diffusion
\begin{equation} \label{eq:dx}
	\mathrm{d}x_s(t) = \mu_s(x_s,t) \mathrm{d}t + \sigma_s(x_s,t) \mathrm{d}W_t,
\end{equation}
with a deterministic part in the form of a drift $\mu_s$, and a stochastic part in the form of a Wiener process, with $\mathrm{d}W_t$ a zero-mean Gaussian white noise. The noise is multiplicative through $\sigma_s$. Both $\mu_s$ and $\sigma_s$ are produced within system $s$. They are structural properties of the system that can change along with the system's structure, with small random variations. Structural changes are assumed to be independent of the noise $\mathrm{d}W_t$. Both are taken to arise from disturbances of the system. Such disturbances may come directly from thermal and quantum noise, and indirectly from long-range electromagnetic and gravitational fluctuations. 

In order to simplify the notation, the subscript $s$ is not written below. Equation~(\ref{eq:dx}) is an It\^o process~\cite{pau13} that becomes another It\^o process when transformed through a function of $x$ and $t$ (It\^o's lemma). For the growth rate $k(x,t)$ this produces
\begin{equation} \label{eq:dk0}
	\mathrm{d}k = \frac{\partial k}{\partial t}\mathrm{d}t + \mu \frac{\partial k}{\partial x}\mathrm{d}t + \frac{1}{2} \sigma^2 \frac{\partial^2 k}{\partial x^2}\mathrm{d}t + \sigma \frac{\partial k}{\partial x} \mathrm{d}W_t.
\end{equation}
Using eq.~(\ref{eq:ki}) and rearranging terms then gives
\begin{equation} \label{eq:dk}
	\mathrm{d}k = \mu \frac{\partial k}{\partial x}\mathrm{d}t + \frac{1}{2} \sigma^2 \frac{\partial^2 k}{\partial x^2}\mathrm{d}t + \sigma \frac{\partial k}{\partial x} \mathrm{d}W_t - \frac{\partial k}{\partial x} \frac{\partial E_t}{\partial t}\mathrm{d}t. 
\end{equation}
The first two terms represent drifts, one produced by $\mu$ and the other produced by the net effect on $k$ of noisy variations along $x$ when $k$ as a function of $x$ is curved ($\partial^2k/\partial x^2 \neq 0$). The last two terms in eq.~(\ref{eq:dk}) are noisy, one produced by the Wiener process and the other by unpredictable changes in the environment. As stated above, if a system is to survive amongst other systems, it should maximize its expected $\mathrm{d}k$ without becoming extinct. Below we will simplify the analysis by taking $\mu = 0$.

The two noisy terms are equally likely positive or negative, with zero mean. Thus maximizing the expected $\mathrm{d}k$ implies maximizing the drift term with $\sigma^2$. However, just maximizing this term through $\sigma^2$ would also increase the noise term depending on $\sigma$. Large noisy variations increase the probability that $\mathrm{d}k$ becomes negative for an extended time, and thereby increase the likelihood that the system's type will become extinct. Therefore, the variance $v_\sigma$ of this noise term needs to be constrained. But it should not be very different from the variance of the last term, $v_E$, which depends on $E_t$ but not on $\sigma$. Making $v_\sigma$ much smaller than $v_E$ would increase the probability of extinction, because then $\sigma$ and thus the drift term would be small, whereas the noise would be nearly constant (almost completely determined by $E_t$). On the other hand, making $v_\sigma$ much larger than $v_E$ would make $E_t$ irrelevant for the dynamics. This would conflict with the basic assumption of the construction here that variations in $E_t$ partly drive the systems' dynamics. 

The relevant time scale for comparing the drift and noise terms is the system's lifetime $\tau$. Through eq.~(\ref{eq:ki}) the growth rate $k$ depends on $z=x-E_t$. The integrals below will be limited to a range $[-Z,Z]$ of $z$ such that beyond this range the partial derivatives of $k$ are sufficiently small to be neglected, that is, $\partial k /\partial z \approx 0$ and $\partial^2 k /\partial z^2 \approx 0$ for $|z|>Z$. Because $E_t$ is assumed to be a random walk-like process, it drifts along the $z$-axis. The range of $z$ it can reach is limited because there is no replication for large $|z|$, but that range is assumed here to be much larger than $[-Z,Z]$. We will therefore assume that the expected values of $z$ produced by $E_t$ in a time $\tau$ are distributed uniformly, at least approximately, over the range $[-Z,Z]$. 

With these simplifying assumptions, constraining the expected noise variance over the system's lifetime $\tau$ requires
\begin{equation} \label{eq:cnstr}
	\frac{\tau}{2Z} \int\limits_{-Z}^{Z}\mathrm{d}z \, \sigma^2 \left(\frac{\partial k}{\partial z}\right)^2 = K,
\end{equation}
where $\langle\mathrm{d}W_t^2\rangle=\mathrm{d}t$ was used~\cite{pau13}, and $K$ is a positive constant such that
\begin{equation} \label{eq:cnstrE}
	K \approx \frac{\sigma^2_E(\tau)}{2Z} \int\limits_{-Z}^{Z}\mathrm{d}z \left(\frac{\partial k}{\partial z}\right)^2 .
\end{equation}
Here $\sigma^2_E(\tau)$ is the expected variance of $E_t$ in a time $\tau$, which depends on the details of $E_t$. Equation~(\ref{eq:cnstrE}) implements the condition discussed above that the noise arising from $E_t$ should neither dominate nor be negligible. However, the precise value of $K$ is not important for the argument below. We can now find the $\sigma(z)$ that maximizes the expected drift in time $\tau$
\begin{equation}
	J = \frac{\tau}{2Z} \int\limits_{-Z}^{Z}\mathrm{d}z \, \frac{1}{2} \sigma^2 \frac{\partial^2 k}{\partial z^2}
\end{equation}
under the constraint of eq.~(\ref{eq:cnstr}). This is an example of an isoperimetric problem that can be solved with the method of Lagrange multipliers~\cite{bru04}. Writing $g(z)=\sigma^2$, $h(z)=\partial k/\partial z$, and $h'(z)=\partial h/\partial z$, then an extremum of $J$ given constraint $K$ implies an extremum of the functional $F$ 
\begin{equation} \label{eq:F}
	F(g,h,h') = \frac{1}{2} g(z) h'(z) - \lambda g(z) h^2(z),
\end{equation}
with $\lambda$ a Lagrange multiplier. Whereas we are interested in finding the function $g$ that maximizes $F$ for a given $h$, we will first find the function $h$ that maximizes $F$ for a given $g$. This will result in a simple, invertible relationship between $g$ and $k$, which subsequently also solves the problem of finding $g$ given $h$. The assumption here is that all functions involved are sufficiently smooth, in particular that $F$ varies smoothly for small variations $\delta h$ and $\delta g$. From the Euler-Lagrange equation
\begin{equation}
	\frac{\mathrm{d}}{\mathrm{d}z}\left(\frac{\partial F}{\partial h'}\right) - \frac{\partial F}{\partial h} = 0
\end{equation}
we find
\begin{equation}
	\frac{\mathrm{d}g(z)}{\mathrm{d}z} + 4 \lambda g(z) h(z) = 0.
\end{equation}
This gives
\begin{equation} \label{eq:gz}
	g(z) = g_0 \mathrm{e}^{-4 \lambda k(z)},
\end{equation}
where $h(z)=\partial k/\partial z$ was used and $g_0$ is a constant. The parameters $g_0$ and $\lambda$ in eq.~(\ref{eq:gz}) can be found numerically from eq.~(\ref{eq:cnstr}). They depend on the detailed form of $k(z)$, which is constrained by eq.~(\ref{eq:dN}). If solutions exist for given parameters, there is a range of possible values $(g_0,\lambda$). The largest value of $\lambda$ gives the largest $J$, because it can be shown that $J=2\lambda K$. This follows from using eq.~(\ref{eq:gz}) for expressing $h$ and $h'$ in terms of $g$ and substituting in the equations for $J$ and $K$. But $\lambda$ cannot be chosen freely, because there is a further constraint on $g=\sigma^2$. The latter is the instantaneous variance of $x$, because eq.~(\ref{eq:dx}) implies $\langle\mathrm{d}x^2\rangle=\sigma^2 \mathrm{d}t$. This variance is not thermal but actively driven, somewhat analogous to that in active matter~\cite{rom12}. Driving the variance consumes a proportional amount of free energy per unit of time. The system must acquire this free energy from its environment. How much is available for varying $x$ depends on the availability of free energy in the environment, on evolved acquisition mechanisms within the system, and on how much free energy the system needs for other processes. We assume here that the result of these factors varies much slower than $x$ and $E_t$, and is effectively independent of them. The rate of available free energy is then effectively a constant that constrains $g(z)$, and thereby $\lambda$.

Quite remarkably, eq.~(\ref{eq:gz}) shows that the $\sigma$ in $\mathrm{d}x$ (eq.~\ref{eq:dx}) that maximizes $\mathrm{d}k$ (eq.~\ref{eq:dk}) is an explicit and very simple function of $k$, with $\sigma^2 \propto 1/\exp(4\lambda k)$. Here $\sigma^2$ only depends on $z$ through $k$ and only depends on $t$ through $z$. Thus the instantaneous variance is inversely related to the instantaneous growth rate. Intuitively, this result can be understood as follows. When the growth rate is larger than zero, the contribution of system $s$ to the population is increasing, and little change in its state is needed. But when the growth rate is smaller than zero, the numbers of system $s$ are declining. If nothing is changed, the system may become extinct. With an increased variance, the state varies faster, which increases the probability that a state with positive growth rate is encountered. If that happens, the variance is decreased automatically, which results in maintained growth, at least until changes in environment or population require further change. Another way to view this mechanism is as a controlled diffusion process. The systems $s$ quickly diffuse away from areas of the state space that have a low growth rate, and much slower away from areas with a high growth rate. In effect, they accumulate in areas with high growth. The efflux from those areas is compensated by a continuous influx of new copies of system $s$ produced by self-replication.

Although the optimal solution is $\sigma^2 \propto 1/\exp(4\lambda k)$, it could not be literally realized in the system. Whereas $\sigma$ is a property of the system (eq.~\ref{eq:dx}), $k$ is the growth rate in eq.~(\ref{eq:dnt}). The growth rate is a non-local variable that is not available to the system in a direct way. The system has no way to measure it directly and instantly. The system can therefore at best approximate $k$ as an internally produced estimate $\hat{k}$. The $\sigma_s$ of eq.~(\ref{eq:dx}) is then a function of $\hat{k}_s$ and not of $k_s$. The estimate $\hat{k}_s$ can gradually evolve and improve in new, random variants of system $s$, because it is advantageous for replication. Only factors to which the system has direct access may be included in $\hat{k}$. For example, the system may get sensors that give information on the state of $E_t$ relative to its own state. Systems that produce a $\hat{k}$ that estimates $k$ better will have a $\sigma^2 \propto 1/\exp(4\lambda \hat{k})$ that is closer to the optimal solution. They will therefore have an expected $\mathrm{d}k$ that is larger than that of other systems. The population will thus gradually become dominated by systems that have adequate $\hat{k}$.

The reason why $\hat{k}$ needs not equal $k$ exactly, is that variations around the optimal $k$ will still produce a near-optimal drift $J$. This follows from the smoothness assumption of the variational approach taken here (eq.~\ref{eq:F} and below). A variation of $\hat{k}$ around the optimum, $k$, produces a variation of $\sigma$ and therefore a variation $\delta g$, which subsequently produces a small change in $F$ and therefore in $J$ as well. Thus $J$ remains close to its optimum. The sensitivity of $\sigma$ to variations in $\hat{k}$ depends on $\lambda$. This is a further reason to constrain $\lambda$, depending on how accurately $\hat{k}$ estimates $k$.

It should be noted that there is no circular logic in the theory developed here. The derivation assumes that eq.~(\ref{eq:dk0}) follows from eq.~(\ref{eq:dx}), and thus that $\sigma$ is not an explicit function of $k$. This assumption seems to conflict with eq.~(\ref{eq:gz}), which has $\sigma$ as a literal function of $k$. But the assumption is correct when taking $\sigma$ as a function of $\hat{k}$. Varying $k$, as in $\mathrm{d}k$, does not affect $\hat{k}$ instantly. Because $\hat{k}$ cannot estimate $k$ with zero lag, $\mathrm{d}\hat{k}$ and $\mathrm{d}k$ are independent locally in time. Therefore, eq.~(\ref{eq:dk0}) still follows from eq.~(\ref{eq:dx}). Estimation with non-zero lag is possible, because $k$ is autocorrelated across many time scales. The latter property follows from eq.~(\ref{eq:ki}) and the fact that $E_t$ is autocorrelated in that way. Also the structural forms of $\hat{k}$ and $\sigma$ cannot change instantly, but only as a result of further evolution of system $s$, with some lag. The actual optimization occurs gradually in real systems. It is therefore cyclical, involving time delays as in a feedback loop, not circular. The theoretical derivation from eq.~(\ref{eq:dk0}) to eq.~(\ref{eq:gz}) just produces a time-averaged short-cut to the ideal end-point of the actual optimization. The result should be seen as an unreachable limit. It seems circular merely because the optimization is static in the theory, whereas it is dynamic and approximate in actual systems.

As an illustration of the theory, we can take $k(z)=k_0 \exp(-z^2/2)-1/\tau$, $\tau=1$, $Z=4$, and $K=1$. In accordance with eq.~(\ref{eq:ki}), this function assumes a maximum growth rate for $z=x-E_t=0$, thus when $x$ matches $E_t$. When the match is poor, for large $|z|$, there is no replication and $n$ declines exponentially. For simplicity, we assume here that the system has evolved a close approximation of $k$. The system thus uses $\sigma(\hat{k})$ with $\hat{k}\approx{k}$. For example, $\hat{k}$ may be based on an approximation of eq.~(\ref{eq:ki}) with $E_{t^-}$ rather than $E_t$, where  $E_{t^-}$ is measured by the system at a time $t^-$ slightly before $t$. The resulting distribution of $n(z)$ depends on the details of $E_t$ and could only be obtained through numerical simulation. In order to get an idea of the order of magnitude of the variables involved, we may assume for this example that $E_t$ is chosen such that $n(z)$ is approximately distributed uniformly in $[-Z,Z]$. Then $\int\mathrm{d}z \, k(z)=0$ (from eq.~\ref{eq:dN}) gives $k_0=3.19$. Solutions of eq.~(\ref{eq:cnstr}) then exist for $g_0$ in the range 0 to 1.43, and $\lambda>0.35$. With $\bar{g}$ the mean of $g(z)$ in $[-Z,Z]$, an energy constraint $\bar{g}=10$ gives $g_0=0.76$ and $\lambda=0.87$, with $J=1.73$, that is, a drift 1.73 times the standard deviation of the noise, $K^{1/2}$. $J$ increases monotonically with $\bar{g}$. Systems that are more effective in harvesting environmental energy therefore have an advantage. Qualitatively similar results were obtained with another functional form for the growth rate, $k(z)=k_0/(1+z^2)-1/\tau$.

The actual $k$ and the estimated $\hat{k}$ have quite different properties with respect to locality. The variable $k$ is a non-local variable of the non-local theory represented by eq.~(\ref{eq:dnt}). The variable is non-local, because it describes the overall effect of a potentially large range of local factors, including stochastic ones. Together these factors produce the growth rate of a system, and they are related to $k$ in an indirect way. But this is not different, in principle, from how the integral form is related to the local form of Maxwell's equations. They are related merely through a well-defined, possibly complex transformation. In contrast, the variable $\hat{k}$ is rather special. Although it is directly defined by strictly local interactions within the system, it produces, in addition, a correlation with $k$. Correlation means here that the zero-lag cross-correlation between $\hat{k}_s(t)$ and $k_s(t)$ is positive, $E[\hat{k}_s(t) k_s(t)]>0$. This correlation is not produced by instantaneous variations of $\hat{k}_s(t)$ and $k_s(t)$, because $\mathrm{d}\hat{k}_s$ and $\mathrm{d}k_s$ are independent. Rather, it is produced by slower changes in $\hat{k}_s(t)$ in response to changes in $k_s(t)$. As stated above, these slower changes are effective because $k_s(t)$ is autocorrelated across many time scales. 

The correlation between $\hat{k}$ and $k$ only exists because system variants with less or no correlation have become extinct. No transformation between $\hat{k}$ and $k$ exists. Yet $\hat{k}$ is effective in maximizing $\mathrm{d}k$ precisely because it has been driven, through competition between different system types, to approximate $k$. In effect, $\hat{k}$ tracks $k$. Part of the causal effectiveness of $\hat{k}$, as promoting system survival, arises from the fact that it tracks $k$. Therefore, the causally effective variable $\hat{k}$ has a non-local scope, through $k$. Equivalently, the non-local variable $k$ thus obtains causal effectiveness that goes beyond that of the local interactions that define $k$. It has obtained causal effectiveness of its own, through $\hat{k}$. It should be noted that there is no conflict with causality here, because non-local spatial effectiveness has to originate from previous $k$, rather than instantaneously. 

\section{Discussion}

Correlation in nature usually arises from direct causal connections or connections with a common cause. Noise generally decreases such correlations over time, although there are exceptions~\cite{gam09}. The theory constructed in the previous section is different on both counts. First, it uses noise to produce rather than destroy correlations. Noise is essential for producing variants with a drift term that utilizes a correlation between $k$ and $\hat{k}$. Second, this correlation does not originate from direct causal connections, but from random generation followed by elimination. Systems with no or little correlation between $k$ and $\hat{k}$ become extinct, leaving the ones that happen to have more correlation, by chance. Crucially, the system dynamics includes multiplicative noise that is coupled to $\hat{k}$, and thereby to the non-local $k$.

The theoretical construction explained above requires a series of assumptions. Although none of these are implausible when taken separately, it is difficult to assess how probable they are in combination. Yet, it should be noted that the goal here was to provide a proof of concept. Counter-intuitively, the theory shows that causal non-locality can indeed arise from local causal interactions. It thereby shows that causal non-locality is possible.

The theory depends critically on the existence of self-replication. Self-replication is rare, but is known to exist in chain reactions of various kinds, in crystal growth, and in autocatalytic chemical processes. But self-replication is most commonly found in biological organisms. Indeed, the theory explained above resembles the Darwinian process of natural selection. Yet, it should be seen as an addition to that process. The regular Darwinian process concerns the factor $\mu(x,t)$ that was deliberately set to zero here. That term produces a drift proportional to $\partial k/\partial x$ (eq.~\ref{eq:dk}). Maximizing this drift requires a $\mu(x,t)$ that at least has the same sign as $\partial k/\partial x$. It would correspond then to a conventional hill climbing optimization. Suitable forms for $\mu(x,t)$ may be found by random variations of systems $s$, as argued by Darwin. However, $\partial k/\partial x$ plays no role in eq.~(\ref{eq:dnt}), not even indirectly. The term $\mu$ can therefore not produce a correlation between a non-local and local variable as the noise term can. Nevertheless, $\mu$ can contribute to non-locality in an indirect way. When the term with $\mu$ in eq.~(\ref{eq:dk}) is positive, the condition on $K$ (eq.~\ref{eq:cnstrE}) can be relaxed, because the system is less vulnerable to downward fluctuations of $\mathrm{d}k$. In addition, the range over which $z$ varies becomes smaller, because $x$ attempts to follow $E_t$. Then $\sigma^2$ can be larger, which increases the drift term that is responsible for producing non-locality.

Biological evolution is obviously much more complex than the mechanisms presented here. In particular, it has a clear separation of the timescales of hereditary change and behavioural change within an organism's lifetime. More complex versions of the model of eq.~(\ref{eq:dx}) that take some of these elaborations into account have been evaluated computationally~\cite{hat15}. Such simulations yield results that are consistent with those derived here more rigorously for a simplified system.

Although the theory presented here is conjectural, it provides a plausible explanation of non-local causality. The correlation between $k$ and $\hat{k}$ is then, presumably, the origin of all more elaborate versions of non-local causality that have subsequently evolved. Examples are the temporal non-locality of memory (genetic, neuronal, and technological), the spatial non-locality of devices such as spider's webs and steam engines, and, probably, even the human ability to produce non-local theories.

\end{document}